\documentclass[10pt,preprint2]{aastex}

\usepackage{bm}
\usepackage{textcomp}
\usepackage{graphicx}
\usepackage{amssymb}
\usepackage{gensymb}

\begin{document}
\title{POLARIZED SCATTERING WITH PASCHEN-BACK EFFECT, HYPERFINE STRUCTURE, AND
PARTIAL FREQUENCY REDISTRIBUTION IN MAGNETIZED STELLAR ATMOSPHERES}
\author{K. SOWMYA$^{1}$, K. N. NAGENDRA$^{1}$, J. O. STENFLO$^{2,3}$
AND M. SAMPOORNA$^{1}$} 
\affil{$^1$Indian Institute of Astrophysics, Koramangala, Bengaluru, India}
\affil{$^2$Institute of Astronomy,
ETH Zurich, CH-8093 Zurich, Switzerland }
\affil{$^3$Istituto Ricerche Solari Locarno, Via Patocchi,
6605 Locarno-Monti, Switzerland}

\email{ksowmya@iiap.res.in; knn@iiap.res.in; stenflo@astro.phys.ethz.ch;
sampoorna@iiap.res.in}
\date{}

\begin{abstract}
$F$-state interference significantly modifies the polarization
produced by scattering processes in the solar atmosphere. Its signature
in the emergent Stokes spectrum in the absence of magnetic fields is
depolarization in the line core. In the present paper, we derive the partial
frequency redistribution (PRD) matrix that includes interference between the upper
hyperfine structure states of a two-level atom in the presence of magnetic fields of
arbitrary strengths. The theory is applied to the Na\,{\sc i} D$_2$
line that is produced by the transition between the lower $J=1/2$ and upper $J=3/2$ states
which split into $F$ states because of the coupling with the nuclear spin $I_s=3/2$.
The properties of the PRD matrix for the single-scattering case is explored,
in particular, the effects of the magnetic field in the Paschen--Back regime and their
usefulness as a tool for the diagnostics of solar magnetic fields.
\end{abstract}


\section{INTRODUCTION}
\label{intro}
The atomic energy levels are split into magnetic substates
in the presence of a magnetic field. When the magnetic splitting is much smaller
than the separation between the hyperfine structure states,
then we are in the Zeeman regime. In this regime, the energy shift produced by
the magnetic field varies linearly with the field strength.
On the other hand, if the magnetic splitting is
comparable to or larger than the energy difference between the
hyperfine structure states, then the magnetic field effects are described by the
Paschen--Back effect (PBE) in which the magnetic splitting varies nonlinearly with
the magnetic field strength, leading to level-crossing interference effects.
The Hanle effect represents a modification of the resonance scattering polarization by the
magnetic field. In the present paper, we are concerned with
the Hanle effect involving hyperfine
structure states. This leads to several interesting phenomena related to level-crossing
interferences.

Hyperfine structure
splitting (HFS) is generally much smaller when compared to fine structure splitting.
Therefore, for those field strengths for which we are still in the Zeeman regime
in the case of fine structure states, we may already be in the regime where PBE is
operating on hyperfine structure states. PBE becomes important in studies of lines
showing hyperfine structure when they are formed in the magnetic
regions on the Sun.

PBE in molecular lines have been extensively studied
both in the context of solar and stellar physics. Molecular PBE
gives signatures in the Stokes profiles, which serve as a promising
tool for diagnosis of solar and stellar magnetic fields \citep[see,
e.g.,][]{ber05,ber06a,ber06,shapetal06,shapetal07,ase06}.
As in the case of molecular lines, PBE also occurs in
atomic lines. The influence of PBE on emergent profiles of
atomic lines such as the He\,{\sc i} 10830 \AA\ multiplet, Fe\,{\sc ii}
multiplet, Si\,{\sc ii}, Si\,{\sc iii} etc., have been studied
\citep[see, e.g.,][]{bom80,sas06,st08,st08a,khal12}.

\citet{landi75} formulated the transfer equation for
a line with hyperfine structure in the presence of a magnetic field, both in LTE
and NLTE. He also presented expressions for
strengths and shifts of the magnetic components of the lines formed
due to transitions between hyperfine structure states. A Fortran program
to compute these strengths and magnetic shifts was made available
in a later paper by the same author \citep[see][]{landi78}. We
use this computer program to calculate the eigenvalues and expansion coefficients
discussed in Section \ref{red-matrix}. \citet{lop02} discussed the net circular
polarization induced by the hyperfine structure and its usefulness
as a tool for the diagnosis of solar magnetic fields in the quiet
photosphere and plages.

The interference between hyperfine structure states
(called the $F$-state interference phenomenon) plays a significant
role in modifying the shapes of the emergent Stokes profiles. \citet{s97}
developed a scattering theory of quantum interference phenomena which
explains the effect of $F$-state interference on coherently
scattered lines. \citet[][hereafter LL04]{ll04} have developed a QED theory to handle
$F$-state interference phenomenon in the PB regime for scattering on
a multi-level atom under the approximation of complete frequency redistribution (CRD).
\citet{cm05} discuss the same problem but for scattering on a multi-term atom
that includes both $J$-state and $F$-state interference phenomena again
under the approximation of CRD. Using the theory of LL04,
\citet{bel07} and \citet{bel09} investigated the effects of magnetic field on
lines resulting from transitions between hyperfine structure states of odd isotope of Ba
and Sc\,{\sc ii}, respectively.

A scattering theory of $F$-state interference based on a metalevel approach was
developed by \citet{landi97}. This theory takes into account PRD in the collisionless
regime. In \citet{smi12}, we
presented the PRD matrix for the $F$-state interference phenomenon
in the absence of magnetic fields and in the collisionless regime.
This PRD theory was applied in \citet{smi13} to illustrate the importance
of PRD, HFS, isotopic shifts, and radiative transfer
in modeling the observed non-magnetic linear polarization profiles of Ba\,{\sc ii} D$_2$
4554 \AA\ line.
In the present paper, we derive the PRD matrix for a two-level atom
with HFS in the presence of a magnetic
field of arbitrary strength.
A straightforward extension of the $J$-state redistribution matrix
(RM) presented in \citet{smi11a} to the case of
$F$-state interference in the PB regime is not possible because
the RM derived in that paper is valid only in the linear
Zeeman regime. Therefore, in the present paper, we formulate the theory
of $F$-state interference in the PB regime and derive an expression
for the RM including PRD in the absence of collisions. We assume the
lower levels to be infinitely sharp and unpolarized. For the sake of
clarity, in Section \ref{atom-sys} we describe the atomic system on which the magnetic
field acts. The atom-radiation interaction in the presence of a
magnetic field of arbitrary strength is discussed quantitatively
in Section \ref{red-matrix}. In Section
\ref{results} we present the characteristics of the RM derived in
Section \ref{red-matrix}. Concluding remarks are presented in Section \ref{conclu}.

\section{THE ATOMIC SYSTEM}
\label{atom-sys}
The atomic system that we consider has two
$J$-states (where $J$ is the total electronic angular momentum)
belonging to two different terms. The lower state is
labeled $J_a$ ($=J_f$, the final state), and the upper state $J_b$.
When the atomic nucleus possesses a spin $I_s$,
the coupling between $J$ and $I_s$ results in hyperfine structure states
$F$ so that ${\bm F}={\bm J}+{\bm I_s}$. The $F$ states are given
by the vector addition formula $F=|J-I_s|,......,J+I_s$. The number of $F$ states
into which a given $J$ state splits is given by $min(2J+1,\ 2I_s+1)$.
For electric dipole transitions
between the $F$ states the selection rules are $\Delta J=0,\pm1$,
$\Delta F=0,\pm1$, and $\Delta \mu=0,\pm1$ in the Zeeman regime and $\Delta J=0,\pm1$
and $\Delta \mu=0,\pm1$ in the PB regime. Here, $\mu$ denotes the magnetic substates
of the hyperfine structure states. The electric dipole nature of the
interaction does not permit transitions among $F$ states of a given $J$ state.

The hyperfine structure of an element has dominant contributions
from magnetic dipole and electric quadrupole interactions \citep[see][]{cor77,woo92}.
The Hamiltonian $\mathcal{H}_D$
describing the interaction of the nuclear magnetic moment with the magnetic field
produced at the nucleus by the valence electrons can be written as
\begin{equation}
\mathcal{H}_D= \mathcal{A}_J {\bm I_s}\cdot{\bm J}\ ,
\label{h1}
\end{equation}
where $\mathcal{A}_J$ is the magnetic dipole hyperfine structure constant
and is mostly determined from experiments. 
The Hamiltonian $\mathcal{H}_Q$ for the electric quadrupole interaction
between the protons and electrons due to the
finite extent of the nuclear charge distribution is given by
\begin{eqnarray}
&& \mathcal{H}_Q=\frac{\mathcal{B}_J}{2I_s(2I_s-1)J(2J-1)}
\nonumber \\ &&
\!\!\!\!\!\!\!\!\!\!\!\!\!\!\!\!\!\!\!\times\bigg\{3({\bm I_s}\cdot{\bm J})^2
+\frac{3}{2}({\bm I_s}\cdot{\bm J})-I_s(I_s+1)J(J+1)\bigg\}\ ,
\label{h2}
\end{eqnarray}
where $\mathcal{B}_J$ is the electric quadrupole hyperfine structure constant
which is also in most cases determined by experimental measurements.

The total Hamiltonian for the atomic system
in the presence of an external magnetic field is written as
\begin{equation}
\mathcal{H}=\mathcal{H}_0+\mathcal{H}_{\rm hfs}+
\mathcal{H}_B\ ,
\label{total-ham} 
\end{equation}
where $\mathcal{H}_0$ is the Hamiltonian describing the atomic structure excluding
hyperfine structure and $\mathcal{H}_{\rm hfs}$ is the Hamiltonian for the hyperfine
structure interaction which, is the sum of $\mathcal{H}_D$ and $\mathcal{H}_Q$.

In the absence of an external magnetic field,
the hyperfine interaction energy $E_{\rm hfs}$ is given by
\begin{eqnarray}
&& E_{\rm hfs} = \frac{1}{2}\mathcal{A}_J \mathcal{K} +
\frac{\mathcal{B}_J}{8I_s(2I_s-1)J(2J-1)}\nonumber \\ &&
\times\{3\mathcal{K}(\mathcal{K}+1)-4J(J+1)I_s(I_s+1)\}\ ,
\label{int-ham}
\end{eqnarray}
where $\mathcal{K}=F(F+1)-J(J+1)-I_s(I_s+1)$.

In the limit of $\mathcal{B}_J\ll\mathcal{A}_J$, the spacing between the $F$ states
is given by the so-called hyperfine structure interval rule
\begin{equation}
\Delta E= E_F - E_{F-1}= \mathcal{A}_J F\ .
\label{f-rule}
\end{equation}
In cases where $\mathcal{B}_J$ is finite, an additional energy shift is
produced.

The magnetic Hamiltonian $\mathcal{H}_B$ in
Equation~(\ref{total-ham}) has the form
\begin{equation}
\ \ \ \ \ \ \ \mathcal{H}_B=\mu_0({\bm L}+2{\bm S})\cdot{\bm B}=\mu_0 B (J_z+S_z)\ ,
\label{mag-ham}
\end{equation}
where $\mu_0$ is the Bohr magneton. The $z$ axis of the coordinate system is
assumed to be along the magnetic field direction. In the PB regime, the magnetic
field produces a splitting comparable with the separation between the $F$ states.
In such cases, the magnetic substates of
a given $F$ state can superpose on the magnetic substates of another
$F$ state. This leads to a mixing of $F$ states. Such a mixing of states
can occur even for very small values of field strengths. The purpose of this
paper is to derive an expression for the PRD matrix representing
the $F$-state interference process in the PB regime.

\section{REDISTRIBUTION MATRIX}
\label{red-matrix}
In the scattering theory approach, the physics of atom radiation interaction is
described through the so-called RM in the astrophysical literature. It describes the
correlations between the incident and scattered photon frequencies, angles, and
polarizations.

\subsection{PB Regime}
\label{pbr}
The PB regime is reached when the Zeeman splitting of the magnetic
substates $\mu$ belonging to a given $F$ state
becomes comparable to the energy
separations between the $F$ states themselves.
This is generally referred to as
the incomplete PB effect. In such a situation,
the magnetic field can no longer be
treated as a perturbation to the atom-radiation interaction,
and one has to carry out a simultaneous
diagonalization of the hyperfine interaction and magnetic Hamiltonians.

The Kramers--Heisenberg formula, which gives the complex probability
amplitudes for scattering from an initial magnetic substate $a$ to
a final substate $f$ through intermediate states $b$, is written as
\begin{equation}
w_{\alpha\beta} \sim \sum_b
\frac{\langle f|{\bf r}\cdot{\bf e}_\alpha|b\rangle
\langle b|{\bf r}\cdot{\bf e}_\beta|a\rangle}{\omega_{bf}-\omega-{\rm i}\gamma/2}\ .
\label{jones}
\end{equation}
Here, $\omega=2\pi\xi$ is the circular frequency of the scattered radiation.
$\hbar\omega_{bf}$ is the energy difference between the excited and final
levels, and $\gamma$ is the damping constant.

The coherency matrix for this scattering process
$a\rightarrow b\rightarrow f$ is given by
\begin{equation}
{\bf W} = \sum_{a}\sum_{f} {\bm w}\otimes {\bm w}^*\ . 
\label{w-mat}
\end{equation}
The incoherent summation is taken over the initial and final levels \citep[see][]{s98}.
Here, ${\bm w}$ is the Jones matrix, and its elements
are given by Equation~(\ref{jones}).

We first identify the basis vectors $|a\rangle, |b\rangle$, and $|f\rangle$
in the PB regime as
\begin{eqnarray}
&& |a\rangle=|J_aI_si_a\mu_a\rangle\ ,
\label{basis-bad}
\end{eqnarray}
with similar forms for $|b\rangle$ and $|f\rangle$.
We then expand these PB regime basis vectors in terms of basis vectors
$|JI_sF\mu\rangle$ of the Zeeman regime as
\begin{eqnarray}
&& \!\!\!\!\!\!\!\!\!\!\!\!\!\!\!\!\!\!
|J_aI_si_a\mu_a\rangle=\sum_{F_a}C^{i_a}_{F_a}(J_aI_s,\mu_a)
\ |J_aI_sF_a\mu_a\rangle\ ,
\label{basis-good}
\end{eqnarray}
with similar expansions carried out for intermediate and final state vectors.
The $C$ coefficients appearing in the above equation are given
by
\begin{equation}
C^i_F(JI_s,\mu)=\langle JI_sF\mu|JI_si\mu\rangle\ ,
\end{equation}
which can be assumed to be real because the total Hamiltonian is real.

Using Equation~(\ref{basis-good}) in the Kramers--Heisenberg
formula and noting that $J_f=J_a$, the dipole matrix elements can be expanded
using the Wigner--Eckart theorem (see Equations. (2.96) and (2.108) of LL04) to obtain
\begin{eqnarray}
&& \!\!\!\!\!\!\!w_{\alpha\beta} \sim \sum_{i_b\mu_b}
\sum_{F_aF_fF_bF_{b^{\prime\prime}}}
\sum_{qq\prime\prime}(-1)^{q-q^{\prime\prime}}  \nonumber \\ && 
\!\!\!\!\!\!\! \times C^{i_f}_{F_f}(J_aI_s,\mu_f)C^{i_a}_{F_a}(J_aI_s,\mu_a)
\nonumber \\ &&\!\!\!\!\!\!\! \times
C^{i_b}_{F_b}(J_bI_s,\mu_b) C^{i_b}_{F_{b^{\prime\prime}}}(J_bI_s,\mu_b)(2J_a+1)
\nonumber \\ && 
\!\!\!\!\!\!\! \times\sqrt{(2F_a+1)(2F_f+1)(2F_b+1)(2F_{b^{\prime\prime}}+1)}
\nonumber \\ && 
\!\!\!\!\!\!\! \times\left (
\begin{array}{ccc}
F_b & F_f & 1\\
-\mu_b & \mu_f & -q \\
\end{array}
\right )
\left (
\begin{array}{ccc}
F_{b^{\prime\prime}} & F_a & 1\\
-\mu_b & \mu_a & -q^{\prime\prime} \\
\end{array}
\right )
\nonumber \\ && 
\!\!\!\!\!\!\! \times\left\lbrace
\begin{array}{ccc}
J_a & J_b & 1\\
F_b & F_f & I_s \\
\end{array}
\right\rbrace
\left\lbrace
\begin{array}{ccc}
J_a & J_b & 1\\
F_{b^{\prime\prime}} & F_a & I_s \\
\end{array}
\right\rbrace
\nonumber \\ && 
\!\!\!\!\!\!\!\times \ 
\varepsilon^{\alpha^*}_{q} \varepsilon^{\beta}_{q^{\prime\prime}}
\ \Phi_{\gamma}(\nu_{i_b\mu_bi_f\mu_f}-\xi)\ .
\label{coh-ele}
\end{eqnarray}
Here, $\varepsilon$ are the spherical vector components of the polarization
unit vectors with $\alpha$ and $\beta$ referring to
the scattered and incident rays, respectively. $\Phi_{\gamma}(\nu_{i_b\mu_bi_f\mu_f}-\xi)$
is the frequency-normalized profile function given by
\begin{equation}
\Phi_{\gamma}(\nu_{i_b\mu_bi_f\mu_f}-\xi)=\frac{1/\pi {\rm i}}
{\nu_{i_b\mu_bi_f\mu_f}-\xi-{\rm i}\gamma/4\pi}\ ,
\label{norm-prof}
\end{equation}
where we have used an abbreviation 
\begin{eqnarray}
&& \!\!\!\!\!\!\!\!\!\!\nu_{i_b\mu_bi_f\mu_f}=\nu_{J_bI_si_b\mu_b,J_aI_si_f\mu_f}
\nonumber \\ &&
\!\!\!\!\!\!\!\!\!\!=\nu_{J_bJ_a} + \frac{E_{i_b}(J_bI_s,\mu_b)
-E_{i_f}(J_aI_s,\mu_f)}{h}\ ,
\end{eqnarray}
with $h$ being the Planck constant. The energy eigenvalues $E$ and the
expansion coefficients $C$ are obtained by diagonalizing the total Hamiltonian
given in Equation~(\ref{total-ham}) \citep[see][]{landi78}. 

We then take the bilinear product of the matrix elements $w_{\alpha\beta}$,
which involves performing coherent summation over the intermediate substates $b$.
Furthermore, we perform incoherent summations over
initial ($a$) and final ($f$) substates to
form the coherency matrix and transform it to the laboratory frame, following
the steps described in Section 2.2 of \citet{sam07a}. With the help of Equation (3.84) of
\citet{s94} and the steps given in Appendix C of \citet{sam07b},
we express the coherency matrix in terms of irreducible spherical tensors
$\mathcal{T}^K_Q(i,\bf n)$ introduced to polarimetry by \citet{landi84}.
Here, $i=0,1,2,3$ refer to the Stokes parameters, $K=0,1,2$, with $Q$ taking ($2K+1$)
values, and $\bm n$ is the direction of the scattered ray.
The coherency matrix is then transformed to Stokes vector basis following the steps in
Appendix C of \citet{sam07b} to obtain
\begin{equation}
S_i=\sum_{j=0}^3 {\bf R}_{ij}^{\rm II}
(x,{\bm n},x^\prime,{\bm n}^\prime;{\bm B})S^\prime_j\ ,
\label{stokes}
\end{equation}
where $S_i$ and $S^\prime_j$ are the Stokes vectors for the scattered
and incident rays, respectively, ${\bf R}^{\rm II}_{ij}$ is the normalized RM for type II
scattering in the laboratory frame and is given by
\begin{eqnarray}
&&\!\!\!\!\!\!\!\!\!\!{\bf R}_{ij}^{\rm II}
(x,{\bm n},x^\prime,{\bm n}^\prime;{\bm B})=
\frac{3(2J_b+1)}{(2I_s+1)}
\nonumber \\ && \!\!\!\!\!\!\!\!\!\! \times
\sum_{KK^\prime Q}
\sum_{i_a\mu_ai_f\mu_fi_b\mu_bi_{b^\prime}\mu_{b^\prime}}
\nonumber \\ && \!\!\!\!\!\!\!\!\!\! \times
\sum_{F_aF_{a^\prime}F_fF_{f^\prime}F_bF_{b^\prime}
F_{b^{\prime\prime}}F_{b^{\prime\prime\prime}}}
\sum_{qq^\prime q^{\prime\prime}
q^{\prime\prime\prime}}
(-1)^{q-q^{\prime\prime\prime}+Q}
\nonumber \\ && \!\!\!\!\!\!\!\!\!\! \times
\sqrt{(2K+1)(2K^\prime+1)}
\cos\beta_{i_{b^\prime}\mu_{b^\prime}i_b\mu_b}
{\rm e}^{{\rm i}\beta_{i_{b^\prime}\mu_{b^\prime}i_b\mu_b}}
\nonumber \\ && \!\!\!\!\!\!\!\!\!\! \times
[(h^{\rm II}_{i_b\mu_b,i_{b^\prime}\mu_{b^\prime}})_{i_a\mu_ai_f\mu_f}+
{\rm i}(f^{\rm II}_{i_b\mu_b,i_{b^\prime}\mu_{b^\prime}})_{i_a\mu_ai_f\mu_f}]
\nonumber \\ && \!\!\!\!\!\!\!\!\!\! \times 
C^{i_f}_{F_f}(J_aI_s,\mu_f) C^{i_a}_{F_a}(J_aI_s,\mu_a)
C^{i_b}_{F_b}(J_bI_s,\mu_b) 
\nonumber \\ && \!\!\!\!\!\!\!\!\!\! \times
C^{i_b}_{F_{b^{\prime\prime}}}(J_bI_s,\mu_b)
C^{i_f}_{F_{f^\prime}}(J_aI_s,\mu_f) C^{i_a}_{F_{a^\prime}}(J_aI_s,\mu_a)
\nonumber \\ && \!\!\!\!\!\!\!\!\!\! \times
C^{i_{b^\prime}}_{F_{b^\prime}}(J_bI_s,\mu_{b^\prime})
C^{i_{b^\prime}}_{F_{b^{\prime\prime\prime}}}(J_bI_s,\mu_{b^\prime})
\nonumber \\ && \!\!\!\!\!\!\!\!\!\! \times
\sqrt{(2F_a+1)(2F_f+1)(2F_{a^\prime}+1)(2F_{f^\prime}+1)}
\nonumber \\ && \!\!\!\!\!\!\!\!\!\! \times
\sqrt{(2F_b+1)(2F_{b^\prime}+1)(2F_{b^{\prime\prime}}+1)
(2F_{b^{\prime\prime\prime}}+1)}
\nonumber \\ && \!\!\!\!\!\!\!\!\!\! \times
\left (
\begin{array}{ccc}
F_b & F_f & 1\\
-\mu_b & \mu_f & -q \\
\end{array}
\right )
\left (
\begin{array}{ccc}
F_{b^\prime} & F_{f^\prime} & 1\\
-\mu_{b^\prime} & \mu_f & -q^{\prime} \\
\end{array}
\right )
\nonumber \\ && \!\!\!\!\!\!\!\!\!\! \times
\left (
\begin{array}{ccc}
F_{b^{\prime\prime}} & F_a & 1\\
-\mu_b & \mu_a & -q^{\prime\prime} \\
\end{array}
\right )
\left (
\begin{array}{ccc}
F_{b^{\prime\prime\prime}} & F_{a^\prime} & 1\\
-\mu_{b^\prime} & \mu_a & -q^{\prime\prime\prime} \\
\end{array}
\right )
\nonumber \\ && \!\!\!\!\!\!\!\!\!\! \times
\left (
\begin{array}{ccc}
1 & 1 & K\\
q & -q^{\prime} & -Q \\
\end{array}
\right )
\left (
\begin{array}{ccc}
1 & 1 & K^\prime\\
q^{\prime\prime\prime} & -q^{\prime\prime} & Q\\
\end{array}
\right )
\nonumber \\ && \!\!\!\!\!\!\!\!\!\! \times
\left\lbrace
\begin{array}{ccc}
J_a & J_b & 1\\
F_b & F_f & I_s \\
\end{array}
\right\rbrace
\left\lbrace
\begin{array}{ccc}
J_a & J_b & 1\\
F_{b^\prime} & F_{f^\prime} & I_s \\
\end{array}
\right\rbrace
\nonumber \\ && \!\!\!\!\!\!\!\!\!\! \times
\left\lbrace
\begin{array}{ccc}
J_a & J_b & 1\\
F_{b^{\prime\prime}} & F_a & I_s \\
\end{array}
\right\rbrace
\left\lbrace
\begin{array}{ccc}
J_a & J_b & 1\\
F_{b^{\prime\prime\prime}} & F_{a^\prime} & I_s \\
\end{array}
\right\rbrace
\nonumber \\ && \!\!\!\!\!\!\!\!\!\! \times
(-1)^Q \mathcal{T}^K_{-Q}(i,{\bm n})
\mathcal{T}^{K^\prime}_Q(j,{\bm n}^\prime)\ .
\label{final-rm}
\end{eqnarray}
Equation~(\ref{final-rm}) represents the PRD matrix for hyperfine interaction
in the PB regime. This equation, when written in the atomic
rest frame, can be directly obtained from Equation~(12) of
\citet{landi97} by introducing the spherical tensors and by assuming that
the lower levels are unpolarized. The PRD matrix derived in this section
satisfies the symmetry relations described in detail
in \citet{bom97b}. In the above expression, the so-called Hanle angle
$\beta_{i_{b^\prime}\mu_{b^\prime}i_b\mu_b}$ is given by
\begin{equation}
 \tan\beta_{i_{b^\prime}\mu_{b^\prime}i_b\mu_b}=
\frac{\nu_{i_{b^\prime}\mu_{b^\prime}i_a\mu_a-}\nu_{i_b\mu_bi_a\mu_a}}{\gamma/2\pi}\ .
\label{hanle-beta}
\end{equation}
The auxiliary functions $h^{\rm II}$ and $f^{\rm II}$ appearing
in Equation~(\ref{final-rm}) have the form
\begin{eqnarray}
&&(h^{\rm II}_{i_b\mu_b,i_{b^\prime}\mu_{b^\prime}})_{i_a\mu_ai_f\mu_f}
\nonumber \\ &&
=\frac{1}{2}[R^{\rm II,H}_{i_b\mu_bi_a\mu_ai_f\mu_f}+
R^{\rm II,H}_{i_{b^\prime}\mu_{b^\prime}i_a\mu_ai_f\mu_f}]\ ,
\label{h-ii}
\end{eqnarray}
and
\begin{eqnarray}
&&(f^{\rm II}_{i_b\mu_b,i_{b^\prime}\mu_{b^\prime}})_{i_a\mu_ai_f\mu_f}
\nonumber \\ &&
=\frac{1}{2}[R^{\rm II,F}_{i_{b^\prime}\mu_{b^\prime}i_a\mu_ai_f\mu_f}
-R^{\rm II,F}_{i_b\mu_bi_a\mu_ai_f\mu_f}]\ ,
\label{f-ii}
\end{eqnarray}
where the magnetic redistribution functions of type II are defined as
\begin{eqnarray}
&& \!\!\!\!\!\!\!\!\!\!R^{\rm II,H}_{i_b\mu_bi_a\mu_ai_f\mu_f}(x_{ba},x_{ba}^\prime,\Theta)
=\frac{1}{\pi\sin\Theta}
\nonumber \\ && \!\!\!\!\!\!\!\!\!\!\times
{\rm exp}\bigg\{-\bigg[\frac{x_{ba}-x_{ba}^\prime+x_{i_a\mu_ai_f\mu_f}}
{2\sin(\Theta/2)}\bigg]^2\bigg\}
\nonumber \\ && \!\!\!\!\!\!\!\!\!\!\times
H\bigg(\frac{a}{\cos(\Theta/2)},\frac{x_{ba}+x_{ba}^\prime+x_{i_a\mu_ai_f\mu_f}}
{2\cos(\Theta/2)}\bigg)\ ,
\label{r-ii-h}
\end{eqnarray}
and
\begin{eqnarray}
&& \!\!\!\!\!\!\!\!\!\!R^{\rm II,F}_{i_b\mu_bi_a\mu_ai_f\mu_f}(x_{ba},x_{ba}^\prime,\Theta)
=\frac{1}{\pi\sin\Theta}
\nonumber \\ && \!\!\!\!\!\!\!\!\!\!\times
{\rm exp}\bigg\{-\bigg[\frac{x_{ba}-x_{ba}^\prime+x_{i_a\mu_ai_f\mu_f}}
{2\sin(\Theta/2)}\bigg]^2\bigg\}
\nonumber \\ && \!\!\!\!\!\!\!\!\!\!\times
2F\bigg(\frac{a}{\cos(\Theta/2)},\frac{x_{ba}+x_{ba}^\prime+x_{i_a\mu_ai_f\mu_f}}
{2\cos(\Theta/2)}\bigg)\ .
\label{r-ii-f}
\end{eqnarray}
Here, $\Theta$ is the scattering angle; the functions $H$ and $F$ are the Voigt and
Faraday--Voigt functions \citep[see Equation (18) of][]{smi11a}.
The quantities appearing in the expressions for the
type II redistribution functions have the following definitions:
\begin{eqnarray}
&& x_{ba}=\frac{\nu_{i_b\mu_bi_a\mu_a}-\nu}{\Delta\nu_{\rm D}}\ ;
\ \ x^\prime_{ba}=\frac{\nu_{i_b\mu_bi_a\mu_a}-\nu^\prime}{\Delta\nu_{\rm D}}\ ,
\nonumber \\ && 
x_{i_a\mu_ai_f\mu_f}=\frac{\nu_{i_a\mu_ai_f\mu_f}}{\Delta\nu_{\rm D}}\ ;
\ \ a=\frac{\gamma}{4\pi\Delta\nu_{\rm D}}\ ,
\label{defn}
\end{eqnarray}
where $x_{ba}$ is the emission frequency, $a$ is the damping parameter,
and $\Delta\nu_{\rm D}$ is the Doppler width.

We remark that the PRD matrix in the PB regime presented in this section can also be
obtained by an alternative approach \citep[see][]{shapetal07} that avoids the use of
statistical tensors $\mathcal{T}^K_Q$.

\subsection{Special Cases}
The PB theory and the relevant RM derived
in Section ~\ref{pbr} gives exact PRD matrix for the
problem at hand. However, it is possible, under limiting
cases, to derive simple expressions for practical
applications. One example of this is the so-called Zeeman regime.
In this regime, the magnetic field is so weak that it
produces a splitting which is much smaller
than the energy differences between the $F$ states.
In such a case, the magnetic Hamiltonian can be
diagonalized analytically using the perturbation theory.

In the Zeeman regime where the basis vector is $|JI_sF\mu\rangle$
in which $F$ is a good quantum number, the RM in Equation~(\ref{final-rm})
takes the form
\begin{eqnarray}
&&\!\!\!\!\!\!\!\!\!\!\!{\bf R}_{ij}^{\rm II}(x,{\bm n},x^\prime,
{\bm n}^\prime;{\bm B}) = 
\frac{3(2J_b+1)}{2I_s+1}\nonumber \\ &&
\!\!\!\!\!\!\!\!\!\!\!\times \sum_{{KK^\prime}Qqq^\prime
{q^{\prime\prime}}{q^{\prime\prime\prime}}F_a\mu_aF_f\mu_f
F_b\mu_bF_{b^\prime}\mu_{b^\prime}} (-1)^{q
-{q^{\prime\prime\prime}}+Q}\nonumber \\ &&
\!\!\!\!\!\!\!\!\!\!\!\times \cos\beta_{F_{b^\prime}\mu_{b^\prime}F_b\mu_b} 
{\rm e}^{{\rm i}\beta_{F_{b^\prime}\mu_{b^\prime}F_b\mu_b}}
\nonumber \\ &&\!\!\!\!\!\!\!\!\!\!\!\times \Big[(h^{\rm II}_{F_b\mu_b,F_{b^\prime}
\mu_{b^\prime}})_{F_a\mu_aF_f\mu_f}
+{\rm i}(f^{\rm II}_{F_b\mu_b,F_{b^\prime}
\mu_{b^\prime}})_{F_a\mu_aF_f\mu_f}\Big]
\nonumber \\ && \!\!\!\!\!\!\!\!\!\!\!\times
 (2F_a+1)(2F_f+1)(2F_b+1)(2F_{b^\prime}+1)\nonumber \\ &&
\!\!\!\!\!\!\!\!\!\!\!\times \sqrt{(2K+1)(2K^{\prime}+1)}\nonumber
\\&&\!\!\!\!\!\!\!\!\!\!\! \times \left\lbrace
\begin{array}{ccc}
J_a & J_b & 1\\
F_b & F_f & I_s \\
\end{array}
\right\rbrace
\left\lbrace 
\begin{array}{ccc}
J_a & J_b & 1\\
F_b & F_a & I_s \\
\end{array}
\right\rbrace \nonumber \\ && \!\!\!\!\!\!\!\!\!\!\!\times
\left\lbrace 
\begin{array}{ccc}
J_a & J_b & 1\\
F_{b^\prime} & F_f & I_s \\
\end{array}
\right\rbrace 
\left\lbrace
\begin{array}{ccc}
J_a & J_b & 1\\
F_{b^\prime} & F_a & I_s \\
\end{array}
\right\rbrace \nonumber \\ &&\!\!\!\!\!\!\!\!\!\!\! \times
\left (
\begin{array}{ccc}
F_b & F_a & 1 \\
-\mu_b & \mu_a & -q^{\prime\prime} \\
\end{array}
\right ) 
\left (
\begin{array}{ccc}
F_b & F_f & 1 \\
-\mu_b & \mu_f & -q \\
\end{array}
\right ) \nonumber \\ && \!\!\!\!\!\!\!\!\!\!\!\times
\left (
\begin{array}{ccc}
F_{b^\prime} & F_a & 1 \\
-\mu_{b^\prime} & \mu_a & -q^{\prime\prime\prime} \\
\end{array}
\right ) 
\left (
\begin{array}{ccc}
F_{b^\prime} & F_f & 1 \\
-\mu_{b^\prime} & \mu_f & -q^{\prime} \\
\end{array}
\right ) \nonumber \\&&\!\!\!\!\!\!\!\!\!\!\!\times
\left (
\begin{array}{ccc}
1 & 1 & K\\
q & -q^{\prime} & -Q \\
\end{array}
\right ) 
\left (
\begin{array}{ccc}
1 & 1 & K^{\prime}\\
q^{\prime\prime\prime} & -q^{\prime\prime} & Q \\
\end{array}
\right )\nonumber \\&&\!\!\!\!\!\!\!\!\!\!\!\times 
(-1)^Q{\mathcal T}^K_{-Q}(i,{\bm n})
{\mathcal T}^{K^\prime}_{Q}(j,{\bm n}^\prime)\ .
\label{red_mat}
\end{eqnarray}
The Hanle angle $\beta_{F_{b^\prime}\mu_{b^\prime}F_b\mu_b}$
is given by
\begin{equation}
 \tan\beta_{F_{b^\prime}\mu_{b^\prime}F_b\mu_b}=
\frac{\omega_{F_{b^\prime}F_b}+
(\mathrm{g}_{F_{b^\prime}}\mu_{b^\prime}-\mathrm{g}_{F_b}\mu_b)\omega_L}{\gamma}\ ,
\label{hanle-angle}
\end{equation}
where $\omega_L$ is the Larmour frequency associated with
the applied magnetic field. The Land\'e factors $\mathrm{g}_F$
appearing in the above equation are defined as
\begin{equation}
\mathrm{g}_F=\mathrm{g}_J\ \frac{1}{2}\frac{F(F+1)+J(J+1)-I_s(I_s+1)}{F(F+1)}\ ,
\label{lande-ji}
\end{equation}
for $F\neq0$. Here, $\mathrm{g}_J$ is the $L-S$ coupling Land\'e factor given by
\begin{equation}
\mathrm{g}_J=1+\frac{1}{2}\frac{J(J+1)-L(L+1)+S(S+1)}{J(J+1)}\ .
\label{lande-ls}
\end{equation}
Equation~(\ref{red_mat}) has a formal
resemblance to the Equation~(25) derived in \citet{smi11a} for
the case of $J$-state interference. Indeed, the $F$-state interference RM in the
Zeeman regime can be obtained from the corresponding $J$-state interference
RM through the replacement of ($J,L,S$) by ($F,J,I_s$) in the latter RM.
When the magnetic field is set to zero in Equation~(\ref{red_mat}),
it takes the same form as Equation~(2) of \citet{smi12}.

\section{SINGLE-SCATTERING REDISTRIBUTION}
\label{results}
To study the behavior of the $F$-state RM derived for arbitrary field
strengths, we consider the atomic line with the following configuration, namely,
the Na\,{\sc i} {\rm D}$_2$ line resulting from the transition between
$J_a=1/2$ and $J_b=3/2$. The wavelength in air corresponding to this transition
is $\lambda_0=5889.95095$ \AA. 
The nuclear spin $I_s=3/2$. The $J--I_s$ coupling results in the hyperfine structure states
$F_b=0,1,2,3$ for the upper state $J_b$ and $F_a=1,2$ for the lower state $J_a$. The
energies of these $F$ states are taken from \citet{ste03}. When the degeneracy of the
magnetic substates of the $F$ states is lifted by the magnetic field, 68 allowed
transitions take place between them in the PB regime. The hyperfine structure
constants have the values $\mathcal{A}_{1/2}=885.81$ MHz, $\mathcal{B}_{1/2}=0$,
$\mathcal{A}_{3/2}=18.534$ MHz, and $\mathcal{B}_{3/2}=2.724$ MHz \citep[see][]{ste03}.
The Einstein {\it A} coefficient for the $J_b=3/2$ state is taken to be
$6.3\times10^7\ {\rm s}^{-1}$. The Doppler width $\Delta\lambda_D=25$ m\AA{} and the
damping parameter $a=0.00227$ (a value obtained after using
$\gamma=6.3\times10^7\ {\rm s}^{-1}$ in Equation~(\ref{defn}), where $a$ is defined)
for all the components. The system that we have considered
obeys the spacing rule described in Section~\ref{atom-sys}.
We study the results of a single 90\textdegree\ scattering event in which
the unpolarized incident beam is scattered by this atomic system in a direction
perpendicular to the incident beam.

\begin{figure}
\begin{center}
 \includegraphics[scale=1.0]{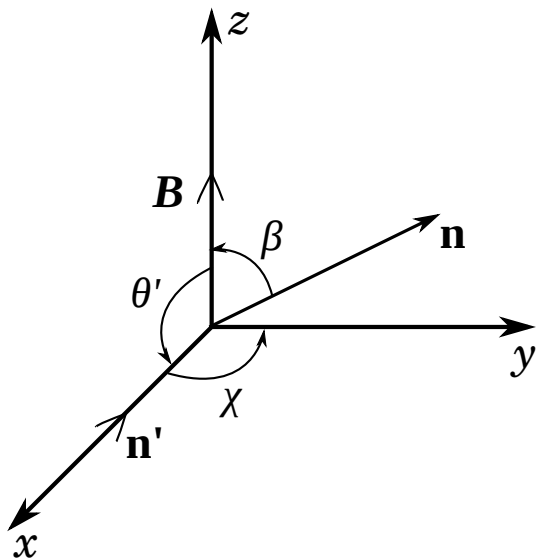}
\end{center}
\caption{Geometry considered for polarization diagrams. $\beta$ is the angle between
the magnetic field vector and the scattered beam. The incident radiation is characterized
by $(\theta^\prime,\chi^\prime)=(90\degree,0\degree)$ and the scattered radiation by
$(\theta,\chi)=(\beta,90\degree)$. The magnetic field inclination $\theta_B=0\degree$
and its azimuth $\chi_B=0\degree$ (magnetic reference frame).}
\label{geometry}
\end{figure}
\begin{figure}
\begin{center}
 \includegraphics[scale=0.6]{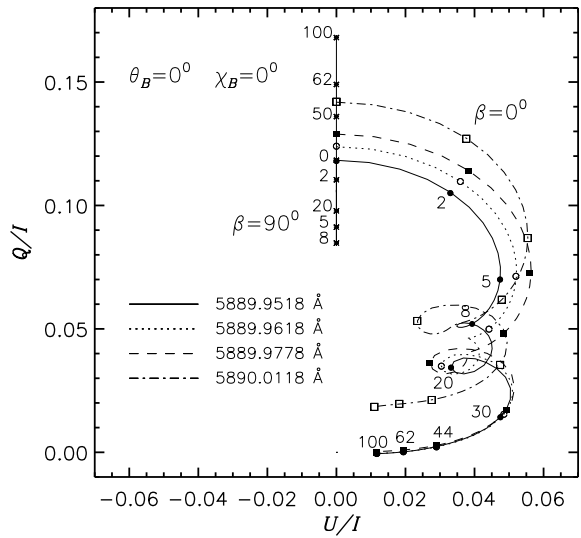}
\end{center}
\caption{Hanle diagrams obtained for $\beta=0\degree$ at different distances from
the line center as indicated in the figure. The solid vertical line represents the
$\beta=90\degree$ case corresponding to the line center wavelength. The numbers along
the solid curves represent the value of magnetic field strength $B$ in G. The symbols on
the other curves mark the same values of $B$ as indicated for the solid curve for
$\beta=0\degree$. The zero field point is the same for the two cases represented by
solid lines.}
\label{pol-diag}
\end{figure}
\begin{table}
 \begin{centering}
\begin{tabular}{cccccc}
\hline
\hline
$F_b\diagdown F_{b^\prime}$ & & 2 & 3 & 3 & 3 \\
\hline
\ \ \ & $\mu_b\diagdown \mu_{b^\prime}$ & $-2$ & $-3$ & $-2$ & $-1$ \\
\hline
1 & 0 & {\bf 12.7} & 31.3 & ... & ... \\

1 & +1 & 15 & 36 & ... & ... \\
\hline
2 & $-1$ & ... & {\bf 25.1} & ... & ... \\

2 & 0 & ... & 16.3 & {\bf 22} & 52 \\

2 & +1 & ... & 14 & 20 & {\bf 44.5} \\

2 & +2 & ... & 13.3 & 18 & 37.8 \\
\hline
\end{tabular}
\caption{Magnetic field strengths (approximate values in G)
for which the magnetic substates
of the $F$-states cross. For instance, the crossing between $\mu_b=0$ of $F_b=1$
and $\mu_{b^\prime}=-2$ of $F_{b^\prime}=2$ occurs at $B\sim12.7$ G. The numbers
highlighted in boldface correspond to the field strength values for which level crossings
occur when one considers the geometry given in Figure~\ref{geometry}, i.e., the
level crossings corresponding to $\Delta\mu=\pm2$.}
\label{tab-1}
\end{centering}
\end{table}

\subsection{Polarization Diagram}
\label{sec2}
The scattering geometry used for the calculation of the polarization
diagrams (plots of $Q/I$ vs $U/I$) in the present section is shown in
Figure~\ref{geometry}.
This geometry is identical to the one considered in Figure~5.11 of LL04.
To explore the effects of the magnetic field in the PB regime on the linear polarization,
we present in Figure~\ref{pol-diag} polarization diagrams 
computed at different distances from the line center.
To construct these diagrams, we first compute the elements of the RM for
a given value of $B$ and integrate the first column of the RM over incident wavelengths.

The solid curve for $\beta=0\degree$
(magnetic field parallel to the scattered beam) matches closely with the Hanle diagram
shown in Figure 10.30 of LL04. As discussed in LL04, the loops seen in the polarization
diagram arise due to the level crossings
that occur in the PB regime. For the atomic system considered here, the level crossing
diagram is identical to that in Figure 3.11 of LL04. The magnetic field strengths
(in G) for which the level crossings occur are given in Table~\ref{tab-1}. For the
geometry considered, level crossings take place only between magnetic substates with
$|\Delta\mu|=2$. The magnetic field values for which these crossings occur are
highlighted in boldface in Table~\ref{tab-1}. The coherence between
the overlapping substates increases around these values of field strengths.
This leads to an increase in the scattering polarization toward its non-magnetic value,
resulting in the formation of loops.

We see from the figure an overall increase in $Q/I$ and $U/I$ as we move away
from the line center when $\beta=0\degree$. Furthermore, the upper loop (near 8 G) seen
in the solid line case disappears for wavelengths away from the line center. On the
other hand, the lower loop (near 20 G) becomes bigger in size. In the far wings of
the line, the polarization diagram
becomes a point corresponding to the Rayleigh case at $Q/I=0.428$ and $U/I=0$.

In Figure~\ref{pol-diag}, we also present the case of $\beta=90\degree$ (magnetic field
perpendicular to the scattering plane). In this case, the Hanle effect in a two-level
atom with HFS shows an interesting phenomenon (see the vertical solid line) called
anti-level crossing, which has been extensively studied and characterized in the case
of CRD \citep[][LL04, p. 604]{bom80}.
We see that the $Q/I$ initially decreases
from 0.118 at $B=0$ to nearly 0.0847 for $B=8$ G. With further increase in $B$,
the $Q/I$ starts increasing and exceeds its value at $B=0$. Thus, we see that
\begin{equation}
 \bigg(\frac{Q}{I}\bigg)^{I_s\neq0}_{B=0}\ <\ \bigg(\frac{Q}
{I}\bigg)^{I_s\neq0}_{B\rightarrow\infty}\ .
\end{equation}
This occurs due to the basis transformation of the energy
eigenstates in the complete PB regime.
The basis transformation takes place when the field strength increases from
incomplete PB regime to the complete PB regime. In the incomplete PB regime, the
energy eigenstates are given by $|JI_si\mu\rangle$, whereas in the complete PB regime
they are given by $|J\mu_JI_s\mu_{I_s}\rangle$. Anti-level crossing is also
known as avoided crossing, in which, due to the strong coupling of the $J$ and $I_s$
to the magnetic field, the magnetic substates instead of crossing, repel each other.
Due to the geometry of the problem, $U/I$ is zero.

\subsection{Stokes Profiles in the PB Regime}
\label{sec3}
In this section, we present the Stokes profiles computed with PRD.
Sections \ref{vertical}, \ref{hori-perp}, and \ref{hori-para} show the Stokes profiles
obtained for various magnetic field configurations. The magnetic field orientations
are discussed in the text and strengths are indicated in the figures.
The incident radiation is characterized by $({\rm cos}\ \theta^\prime,\chi^\prime)=
(0,0\degree)$ and the scattered ray by $({\rm cos}\ \theta,\chi)=(0,90\degree)$.
For the computation of the Stokes profiles, we use a wavelength grid
having 376 finely spaced points covering a bandwidth of 2 \AA. 
The separation between the $F$ states in the absence of a magnetic field is of the
order of m\AA. In the presence of a magnetic field, the magnetic components
are shifted away from the line center and the wavelength grid that we have
considered is good enough and covers all the components shifted by the magnetic field.

\begin{figure*}
\begin{center}
 \includegraphics[scale=0.5]{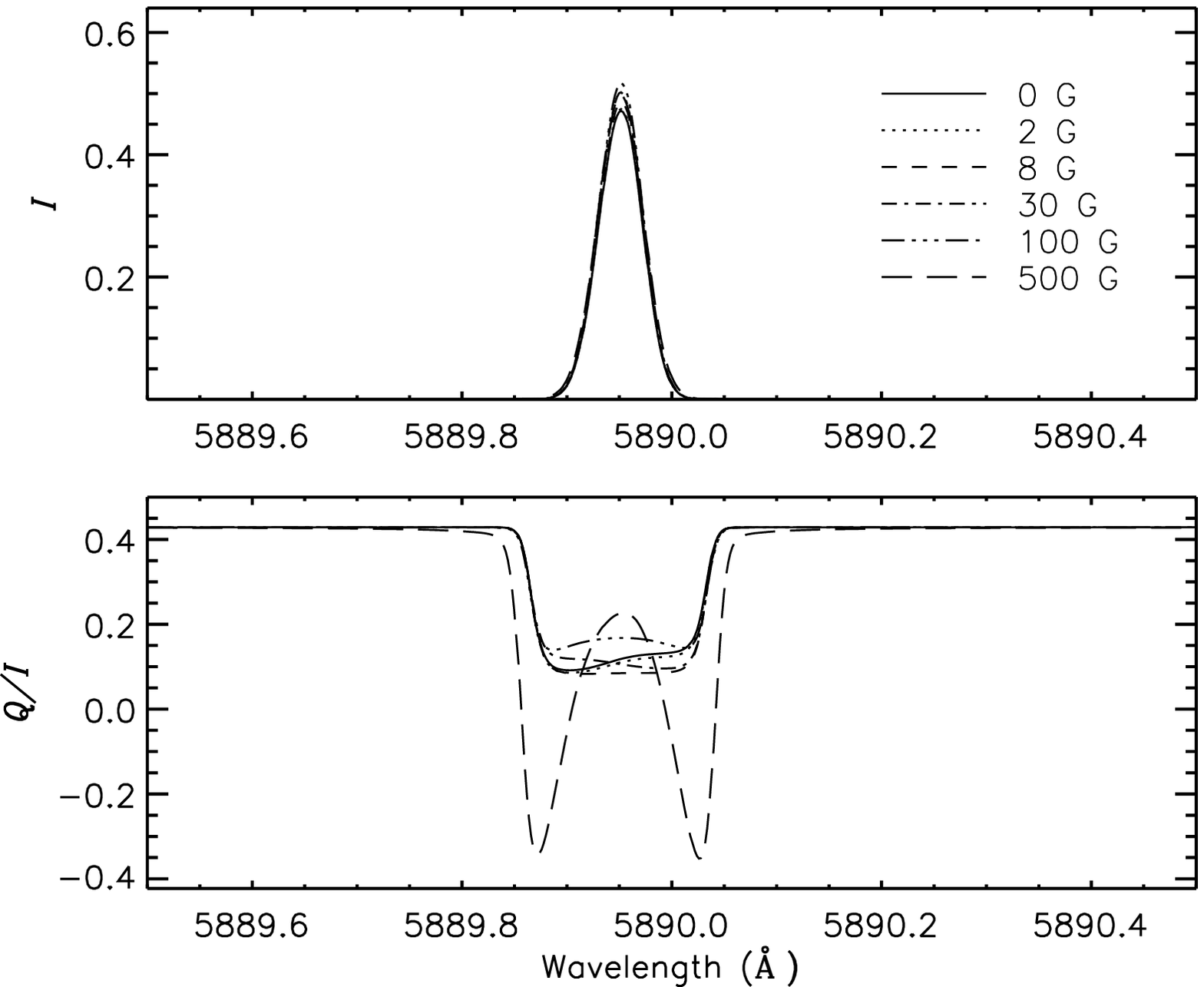}
 \includegraphics[scale=0.5]{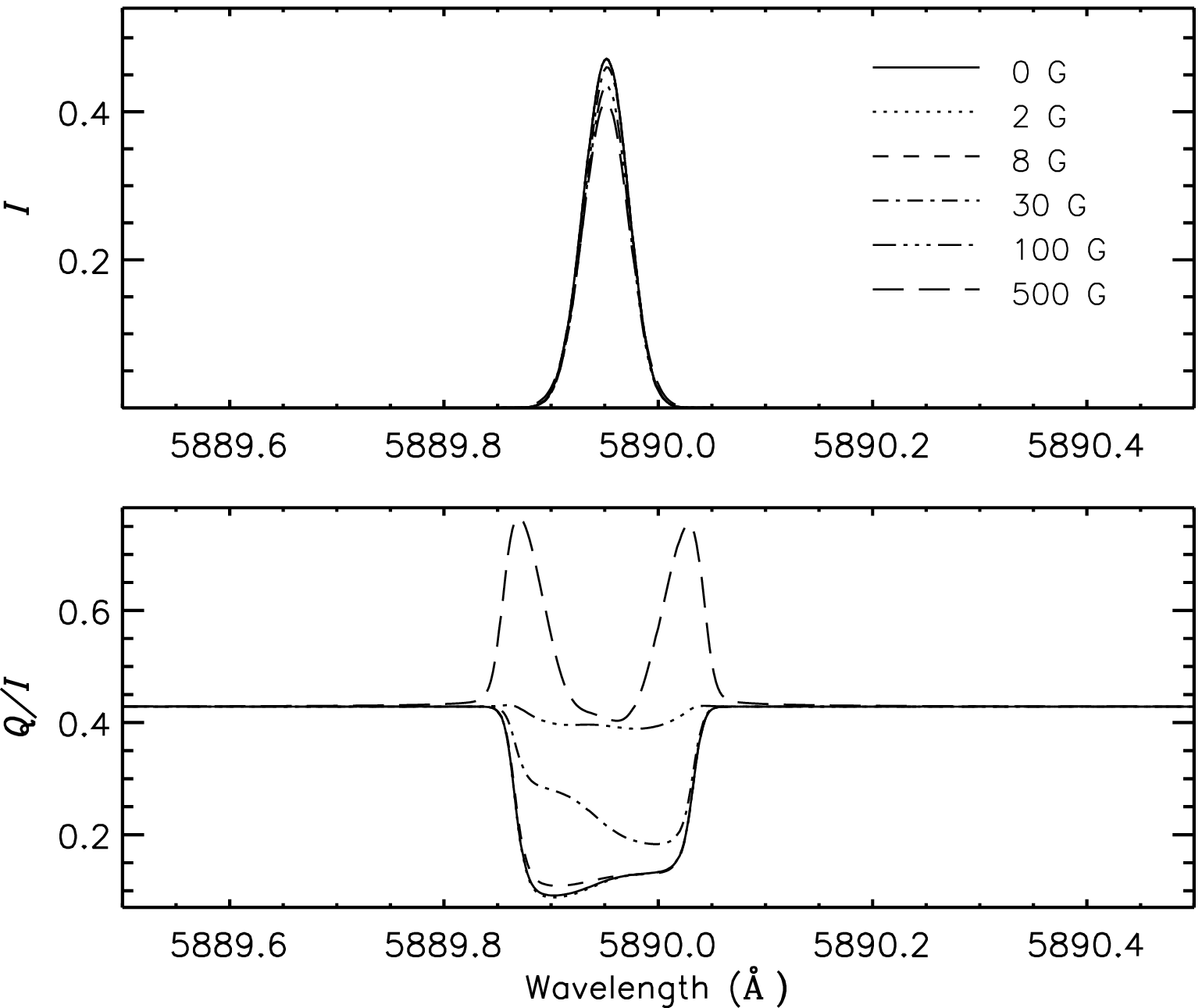}
\end{center}
\caption{Stokes profiles computed for the case of a vertical magnetic
field (left panels) and for the case of a horizontal magnetic field
perpendicular to the line of sight (right panels).}
\label{vert}
\end{figure*}
\begin{figure*}
\begin{center}
 \includegraphics[scale=0.45]{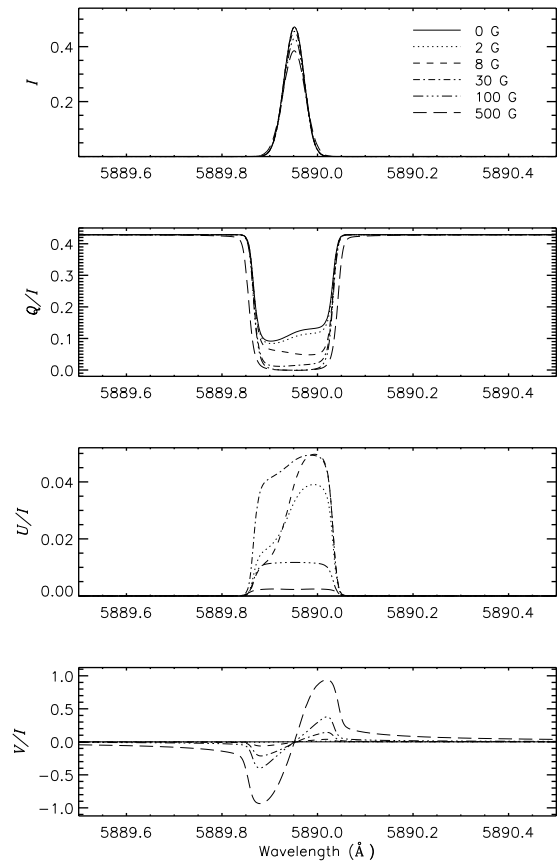}
\end{center}
\caption{Stokes profiles computed for the case of a horizontal magnetic field
parallel to the line of sight.}
\label{hor-pa}
\end{figure*}

\subsubsection{Vertical Magnetic Field Perpendicular to the Line of Sight}
\label{vertical}
In Figure~\ref{vert}, the left panels show the Stokes profiles obtained for different
strengths of a vertical magnetic field ($\theta_B=0\degree$ and $\chi_B=0\degree$). 
We see that the intensity increases slightly with increasing field strength.
$Q/I$ profiles show a decrease in amplitude up to 8 G (see short dashed line). For
stronger fields (greater than 8 G), the $Q/I$ amplitude increases (see also
Figure~\ref{pol-diag}). This is the signature of anti-level crossing effect which
occurs because of the repulsion between the magnetic substates. As discussed earlier,
as a result of this effect, the $Q/I$ line core value, when considered as a function of
field strength, initially decreases and then increases beyond its non-magnetic value.
Transverse Zeeman effect signatures show up prominently for fields stronger than 100 G.
Because of the geometry considered, $U/I$ is zero.

\subsubsection{Horizontal Magnetic Field Perpendicular to the Line of Sight}
\label{hori-perp}
The case of a horizontal field perpendicular to the line of sight
($\theta_B=90\degree$ and $\chi_B=0\degree$) is shown in the right panels of
Figure \ref{vert}. We see that the intensity decreases monotonically with
field strength. $Q/I$ profiles show an increase in amplitude from their Rayleigh
scattered values with an increase in the field strength. For fields of the order of
100 G and larger, we see three lobed profiles in $Q/I$ due to transverse Zeeman effect.
Once again, $U/I$ is zero due to the geometry.

\subsubsection{Horizontal Magnetic Field Parallel to the Line of Sight}
\label{hori-para}
For this geometry of the magnetic field ($\theta_B=90\degree$ and $\chi_B=90\degree$),
the intensity profiles behave in the same way as in the case of a horizontal field
perpendicular to the line of sight (see Figure \ref{hor-pa}). The depolarization in the
line core due to the Hanle effect is clearly
visible in the $Q/I$ panel. The $U/I$ signal is now
generated because of Hanle rotation. There is an increase in $U/I$ amplitude for weaker
fields and then a decrease for stronger fields, which is a typical signature of the Hanle
effect. We notice that the $V/I$ profiles are asymmetric in the incomplete PB regime
(for fields up to 200 G) while it is perfectly anti-symmetric in the complete PB
regime (for fields greater than 200 G). This is because incomplete
PBE shifts the magnetic components asymmetrically about the line center and causes
differential strengths for these components. Because of this asymmetry, the net circular
polarization (NCP), defined as $\int V\ d\lambda$ (where the integration is done over
the full line profile), is non-zero (the NCP would be zero if the splitting produced by
the magnetic field is symmetric). For the atomic line under consideration, NCP remains
non-zero up to 200 G.

\section{CONCLUSIONS}
\label{conclu}
LL04 incorporated the PBE on the hyperfine structure states in the polarization
studies under the assumption of CRD. They assume that the incident radiation is
independent of frequency in an interval larger than the frequency shifts and inverse
lifetimes of the hyperfine structure substates involved in the transitions (flat-spectrum
approximation). In the present paper, we have considered the same problem, but for the
case of PRD. This allows us to handle the frequency dependence of the incident radiation
field (relaxation of flat-spectrum approximation). In this way, the Stokes profile shapes
can be properly calculated by including the effects of PRD. We have derived the
PRD matrix for $F$-state interference process, in the collisionless regime, in the
presence of magnetic fields of arbitrary strengths.

Through the polarization diagrams computed at different scattered wavelengths,
we have shown the dependence on wavelength of the loops which are characteristics of
level crossings that occur in the PB regime. With the help of the Stokes profiles
computed for the case of a vertical magnetic field, we have also demonstrated the
anti-level crossing effect, which was discussed for the case
of CRD by \citet{bom80} and LL04. Based on the formulation described in the
present paper, it is possible to use the diagnostic potential of PBE with PRD, in a
complementary way with the Zeeman effect, to determine the strengths and geometry of
the magnetic fields in the solar atmosphere.

We thank Ms. H. N. Smitha for her help and useful discussions. We also acknowledge
the use of HYDRA cluster at the Indian Institute of Astrophysics for computing.
We are grateful to the referee for very useful comments and suggestions which helped
to improve the paper substantially.

\label{lastpage}
\end{document}